\documentclass[a4paper,11pt]{article}
\usepackage{pos}

\author{Frederik Hermann Lauber (for the IceCube Colloboration)}

\FullConference{37$^{\rm{th}}$ International Cosmic Ray Conference (ICRC 2021)\\
		July 12th -- 23rd, 2021\\
		Online -- Berlin, Germany}

\title{New Flux Limits in the Low Relativistic Regime for Magnetic Monopoles at IceCube}
\ShortTitle{New Flux Limits for Low Relativistic Magnetic Monopoles}
\emailAdd{flauber@icecube.wisc.edu}

\abstract{
	Magnetic monopoles are hypothetical particles that carry magnetic charge. Depending on their velocity, different light production mechanisms exist to facilitate detection. In this work, a previously unused light production mechanism, luminescence of ice, is introduced. This light production mechanism is nearly independent of the velocity of the incident magnetic monopole and becomes the only viable light production mechanism in the low relativistic regime (0.1-0.55c). An analysis in the low relativistic regime searching for magnetic monopoles in seven years of IceCube data is presented. While no magnetic monopole detection can be claimed, a new flux limit in the low relativistic regime is presented, superseding the previous best flux limit by 2 orders of magnitude.

}

\usepackage[ragged]{sidecap}
\usepackage{cleveref}

\begin{document}
\maketitle
\section{Introduction}
The IceCube Neutrino Observatory, IceCube for short, is a cubic-kilometer neutrino detector installed in the ice at the geographic South Pole \cite{Aartsen:2016nxy}  between depths of 1450\,m and 2450\,m, completed in 2010.
While designed to detect neutrinos, it can also be utilized to detect any particles passing through the ice while producing light.
5160 Digital Optical Modules, DOMs for short, record light pulses inside the ice. 

In this contribution, we utilize luminescence light for the first time at IceCube, to explore for hypothetical magnetic monopoles in the low relativistic ( 0.1c  to 0.55) regime.
This complements past searches at IceCube that have utilized Cherenkov signals from potential monopole induced proton decay~\cite{Aartsen2014}, and direct and indirect Cherenkov light from monopoles~\cite{1511.01350v2} as light production channels.
A cut and count based analysis optimized on simulated events and evaluated on seven years of IceCube data is presented.

\section{Magnetic Monopoles}
While no magnetic monopoles have been observed up to now, certain properties of magnetic monopoles can be derived independently from other assumptions.
Their magnetic charge has to be a multiple of the Dirac charge, $g_D$~\cite{Dirac:423077}, defined as
\begin{equation}
	\label{eq:dirac:val:approx}
 	g_D  = \frac{\hbar \textrm{c}}{2 q_e} = \frac{q_e}{2 \alpha} \approx 68.5 q_e.
\end{equation}

\begin{SCfigure}
    \includegraphics[width=0.5\textwidth]{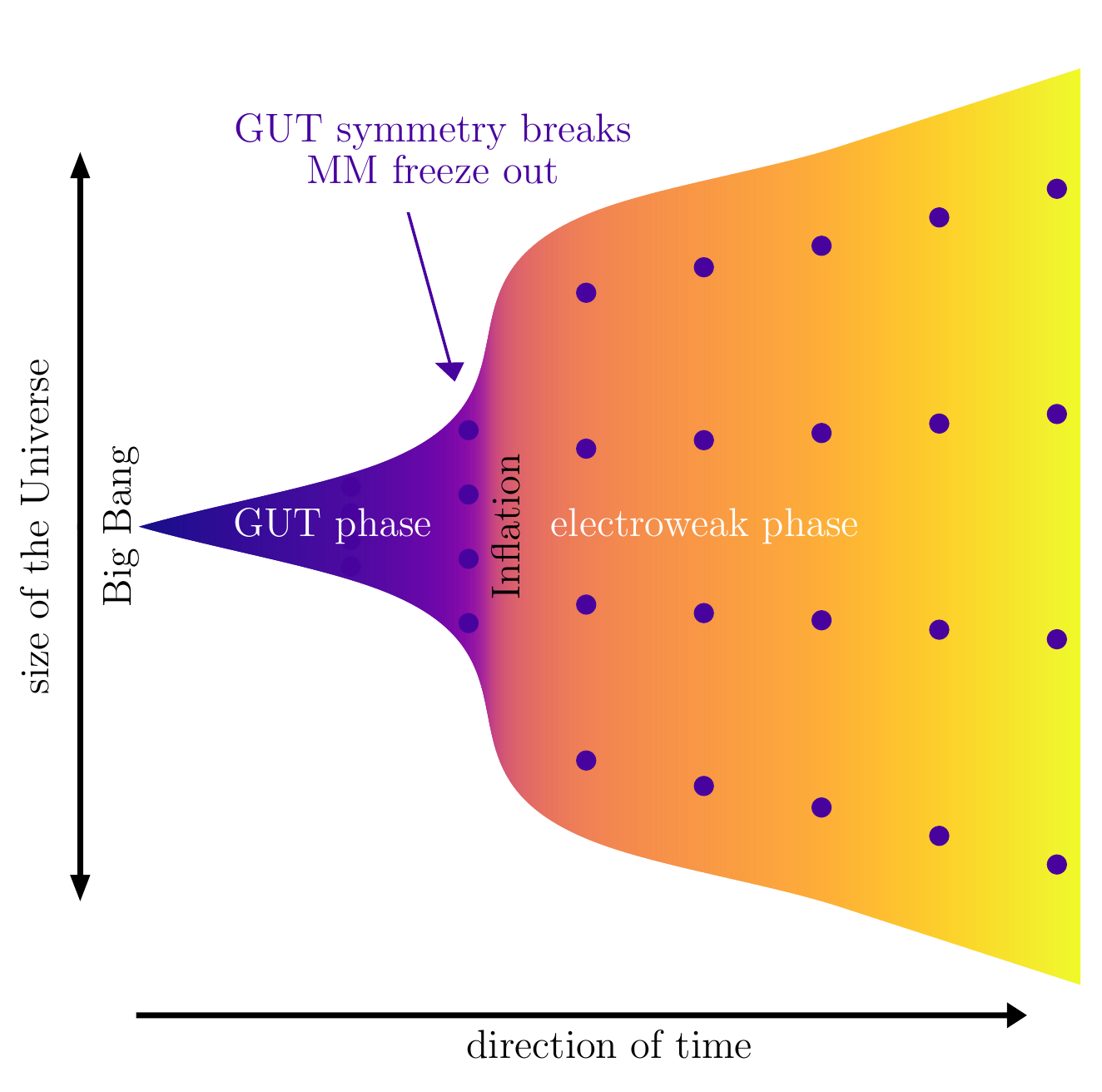}
	\caption{
	A sketch of the freeze out of relic magnetic monopoles during the early epochs of the Universe is shown.
	The energy density of the Universe is indicated by a color shift from purple to yellow.
	The Universe starts out in a high energy density state, also called the GUT-phase.
	With the expansion of the Universe, the energy density decreases until the GUT symmetry spontaneously breaks and magnetic monopole freeze out occurs. 
	The remaining monopoles maintain their internal energy density and are diluted by the following inflation of the Universe.
	}
	\label{fig:gut_freeze}
\end{SCfigure}
\begin{figure}
\centering
\begin{minipage}{0.49\textwidth}\centering
\includegraphics[width=\textwidth]{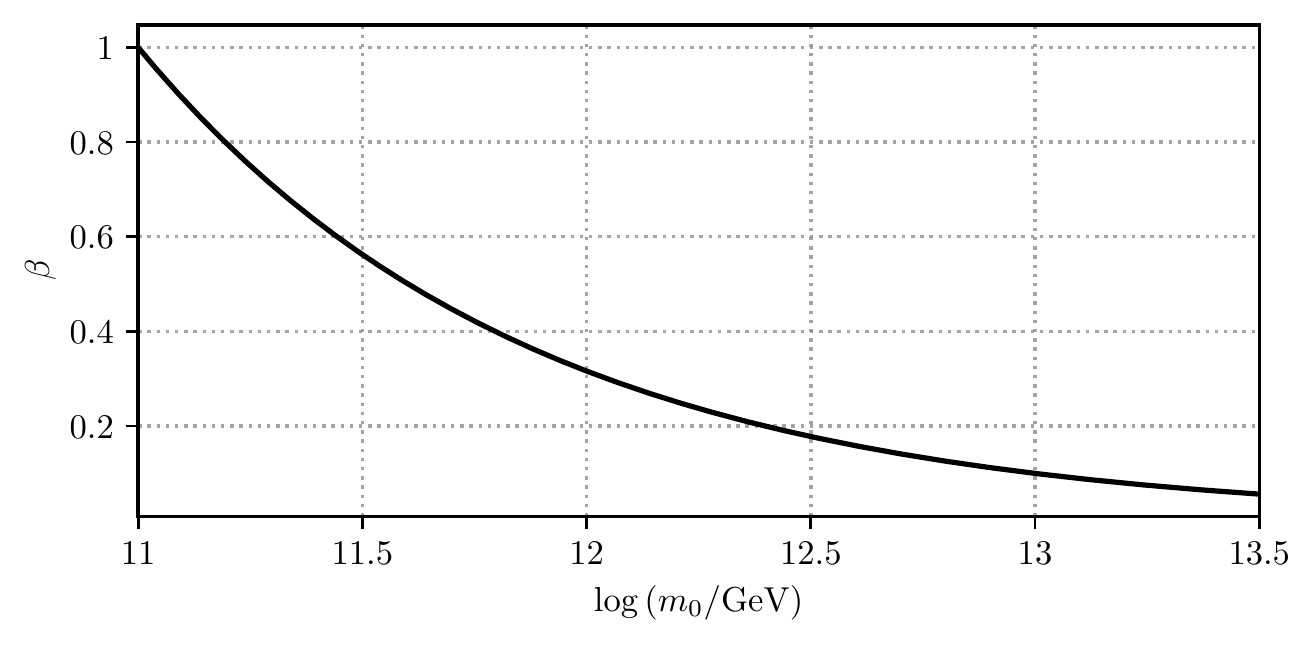}
\caption{The expected velocity of a magnetic monopole with a Dirac magnetic charge at the position of Earth as a function of the rest mass is drawn assuming acceleration only inside the Milky Way~\cite{Tanabashi:2636832}.}
\label{fig:imms_velo}
\end{minipage}\hfill
\begin{minipage}{0.49\textwidth}
\centering
\includegraphics[width=\textwidth]{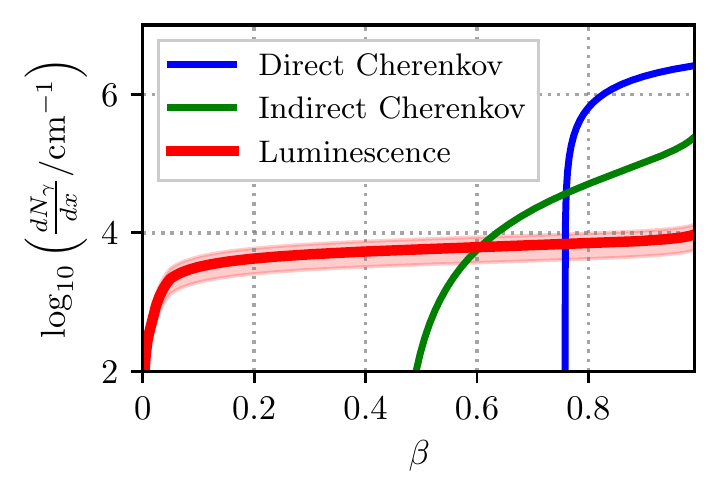}
\caption{The number of expected photons per unit length as a function of velocity for different light production channels is depicted.
The shaded region around the solid luminescence line indicates the region covered by a $\pm40\%$ systematic shift of the effective luminescence light yield. }
\label{fig:light_types}
\end{minipage}
\end{figure}

Relic magnetic monopoles are assumed to have been created during the early epochs of the Universe as depicted in \cref{fig:gut_freeze}.
While they would be created with negligible kinetic energy, they undergo acceleration by cosmic magnetic fields.
Depending on the rest mass and magnetic charge of the magnetic monopole and the size and coherent length of cosmic magnetic fields, their expected velocity differs greatly.
For example, assuming acceleration within the Milky Way galaxy from relative rest for a Dirac charged magnetic monopole, a shift from velocities close to the speed of light to non-relativistic velocities can be observed in the mass range of $m_0=10^{11}\,\textrm{GeV to }10^{13.5}\,\textrm{GeV}$ as depicted in \cref{fig:imms_velo}.

Different light production channels exist to detect magnetic monopoles passing through ice, each dominating at different velocities.
Starting close to the speed of light, magnetic monopoles can induce Cherenkov light just like any other highly electrically charged particle.
Below the Cherenkov threshold in ice ($\approx 0.76\,\textrm{c}$), no 
direct Cherenkov light is produced anymore.
Instead, indirect Cherenkov light, Cherenkov light produced by secondary $\delta$-electrons induced by a passing magnetic monopole, becomes the dominant light production mechanism down to about 0.6\,c.
In contrast to direct Cherenkov light, there is no sharp cut-off. 
Instead, the light yield decreases until luminescence becomes the dominant light production mechanism.
Luminescence light is mostly velocity independent but has a lower overall light yield in contrast to the two other aforementioned light production channels.
A comparison of the three described light production channels is depicted in \cref{fig:light_types}.

Luminescence light is induced by the energy transferred between the passing magnetic monopole and the surrounding ice.
In the velocity regime of 0.05\,c to 0.99995\,c, magnetic monopoles passing through matter lose kinetic energy dominantly by excitation and ionization of electrons in the target material.
This is modeled by the magnetic charge adjusted Bethe-Bloch formula~\cite{ahlen1978stopping}.
Parts of this transferred energy can be turned into detectable light~\cite{Quickenden_1982,Freeman_1984,ScienceDirect:S016943320801369X}.
This transference is dependent on the temperature and impurities of the ice~~\cite{quickenden1991effect,ScienceDirect:135901978890166X}.
Dedicated in-situ measurements of the luminescence of the ice utilized by IceCube has been conducted and an effective luminescence light yield in the effective wavelength ranges detectable by IceCube of $\frac{dN_\gamma}{dE}=1 \frac{\textrm{photon}}{\textrm{GeV}}$ has been measured~\cite{pollmann2019enabling}.
The product of the effective luminescence light yield and the kinetic energy loss is the aforementioned luminescence light yield. 

\section{Simulation}
To allow for an unbiased selection process, all selections are based on simulated background and signal events.
Simulated background events are then compared to a statistically blinded set of measured events to validate the background simulation.
As background simulation, a combination of CORSIKA~\cite{heck1998corsika} based cosmic ray induced air shower simulation weighted to a GaisserH3a~\cite{gaisser2011spectrum} flux model and atmospheric electron and muon neutrinos, weighted to a HKKM2006~\cite{astro-ph/0611418v3} flux, is used.

For signal simulation, a flat velocity spectrum between 0.1\,c and 0.6\,c has been chosen, as neither a a-priori assumed spectrum of magnetic monopoles exists nor is processing time strongly dependent on the velocity.
However, two luminescence light yields are utilized: $\frac{dN_\gamma}{dE}=1 \frac{\textrm{photon}}{\textrm{GeV}}$ as the measured, most likely scenario for IceCube ice and $\frac{dN_\gamma}{dE}=0.2 \frac{\textrm{photon}}{\textrm{GeV}}$ as a worst case assumption to harden selection steps against possible systematic shifts of the effective luminescence light yield.  

\section{Event Selection}
Magnetic monopoles in the low relativistic regime would have the form of slow, track like particles passing through the ice utilized by IceCube.
Additionally, light should be emitted homogeneously around the track as energy is also deposited homogeneously in the ice.
Thus, a long, time- and space-like, track with few gaps passing through the whole of the fiducial volume of IceCube is ideal.
Only downward-going events are regarded as the Earth can potentially shield IceCube from magnetic monopoles with higher magnetic charges than the Dirac charge.

The candidate event selection is conducted in three stages.
First, an event needs to pass the standard triggers of IceCube~\cite{Aartsen:2016nxy} which select 30\,\% to 50\,\% of all simulated signal events.
Next, a set of requirements are added to reduce the candidate event rate down to 0.6\,Hz, while keeping between 30\,\% and 90\,\% of the simulated signal depending on the velocity of the incident magnetic monopole before the final selection step, described in the next section, is applied.

At least 25\,DOMs must have detected a light pulse.
The time between all first light pulses at the DOMs must be at least 4000\,ns.
Afterwards, a track hypothesis~\cite{aartsen2014improvement} is calculated which must converge and reconstruct a velocity between 0.1\,c and 0.6\,c.
For the next steps, only light pulses in a 100\,m radius around the track hypothesis are regarded.
The position of the light pulses are projected onto the track hypothesis.
The distance of the center of gravity of the projected positions of the light pulses of the time sorted first and last quartile of light pulses must be at least 250\,m.
Additionally, the maximal distance between two projected hits on the track hypothesis must be below 200\,m.
The event is split time-wise in two parts which are used to create individual track hypothesis.
The first track hypothesis's velocity must be between 0.15\,c and 0.65\,c while 
the second track hypothesis's velocity is required to be between 0.07\,c and 0.8\,c.
The softer requirement on the second track hypothesis's velocity is due to this selection step being applied after the first one as well as the first one having a better reconstruction quality due to the usage of early, more reliable light in the first place.
As a last step, events passing through corners of the fiducial volume are removed.
The fiducial volume is modeled by a cylinder with radius and height of 750\,m at the center of IceCube.
The initial track hypothesis has to pass through at least 250\,m of this cylinder for the event to be regarded as a possible event.

\section{Final Machine Learning Based Selection}
A machine learning based algorithm is applied based on the XGBoost framework~\cite{Chen:2016:XST:2939672.2939785} to make the final selection of candidate events.
For each event, $24$ features are identified to separate simulated background events from simulated signal events while disfavoring separation between simulated background and the statistically blinded, measured data.

As background simulation is statistically limited, a bootstrap aggregating~\cite{breiman1996bagging} based approach is chosen.
$1000$ Boosted Decision Trees, BDTs for short, are trained on randomly sampled subsets of the available training data.
Each event has a 10\,\% chance to be included in each subset.
The classification of a BDT on an event which was used to train the BDT is discarded.
Thus, for each event, there are on average $900$ classifications mapping the event to score $s_i\in \left[0, 1\right]$ where $1$ is signal-like and $0$ is background-like.

\begin{SCfigure}
    \includegraphics[width=0.5\textwidth]{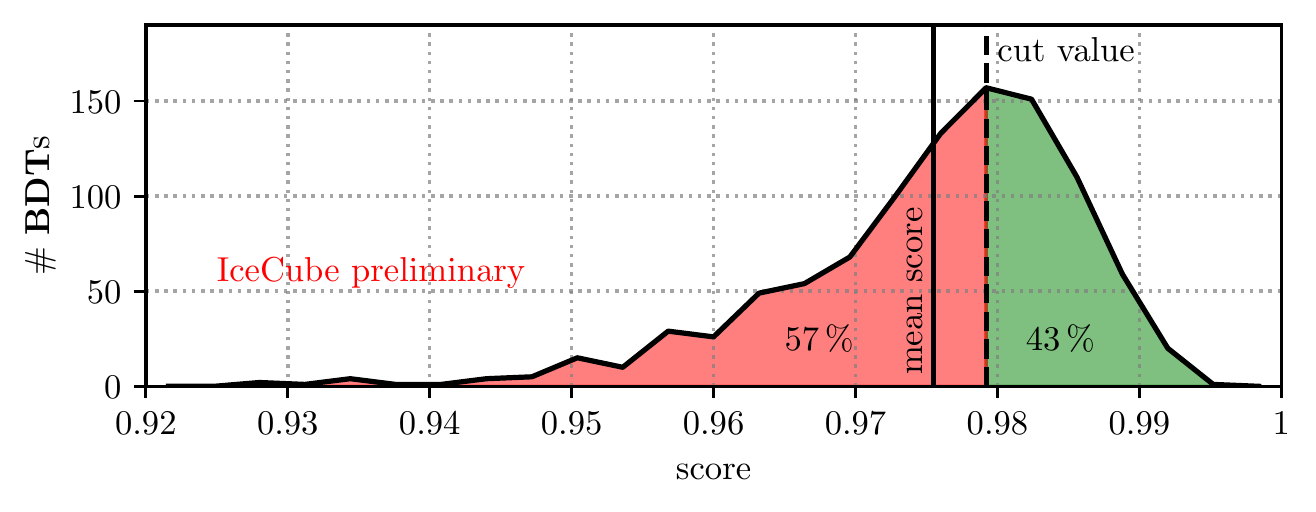}
\caption{The number of BDTs predicting a specific score for a single, randomly selected event is shown.
Assuming an exemplary selection cut at about 0.979, this event would be rejected based on the mean predicted score.
Alternatively, the event can contribute by a reduced weight, here only 43\,\% to enhance the statistics of projected distributions after the selection step.
}
\label{fig:bagging_examle}
\end{SCfigure}

This set of classifications can be interpreted as the probability density for the event to pass the classification process.
By allowing events to partially pass the event selection, statistically limited distributions after the final selection step can be estimated.
A sketch of the setup is depicted in \cref{fig:bagging_examle}.

To select the final cut value $C$ so $\hat{s}\geq C$ is indicative of the final sample of candidate events, the Model Rejection Factor, MRF for short,
is defined as
\begin{equation}
	\textrm{MRF} \left(c\right) = 
	\frac{
		\mu_N^{90\,\%} \left(c\right)
	}{
		N_S \left(c\right)
	}
\end{equation} 
where $N_S$ is the number of expected remaining signal events and 
$\mu_N^{90\,\%} \left(\textrm{c}\right)$ is the upper number of true signal counts compatible
with the observed counts and the predicted average number of background events if the experiment was repeated an infinite amount of times in 90\,\% of the cases.

\begin{figure}
\centering
\begin{minipage}{0.49\textwidth}
    \includegraphics[width=\textwidth]{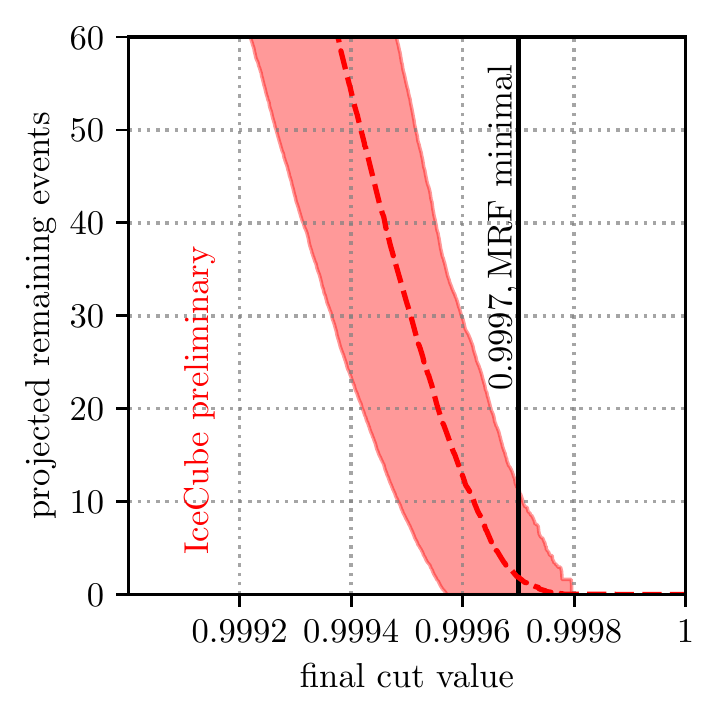}
\caption{The number of remaining background events as a function of the final cut value is illustrated.
The dashed line indicates the mean value while the contour indicates the upper and lower number of events expected.
At 0.9997, a vertical line is drawn indicating the location at which the MRF becomes minimal.}
\label{fig:mrf}
\end{minipage}\hfill
\begin{minipage}{0.49\textwidth}
\centering
    \includegraphics[width=\textwidth]{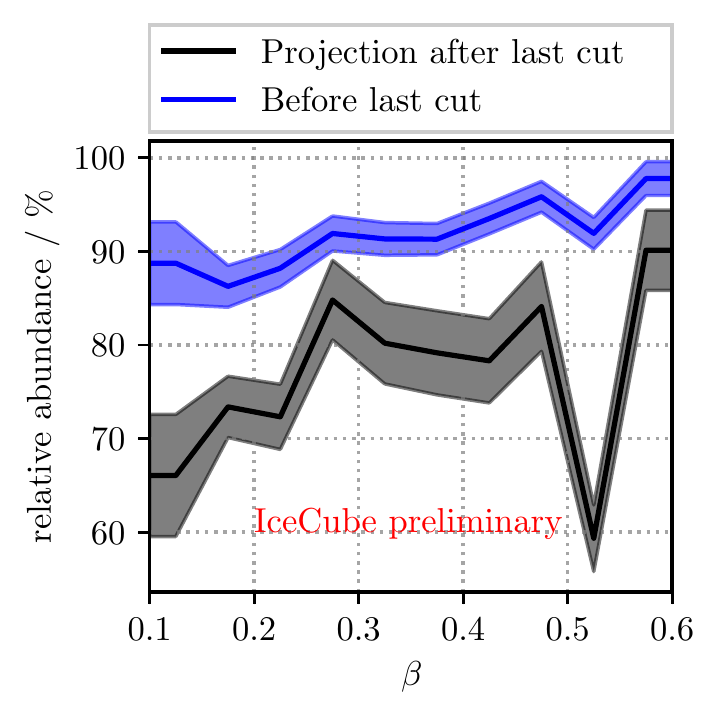}
\caption{The relative abundance of signal events due to systematic shifts as a function of the velocity is drawn.
The black solid line indicates the mean expected shift at the final event selection stage while the the {\color{blue}blue} solid line is the expected shift right before the final selection step. The contours indicate the statistical uncertainties.}
\label{fig:sys}
\end{minipage}
\end{figure}

$\mu_N^{90\,\%} \left(\textrm{c}\right)$ is calculated based on 
Feldman-Cousin approach~\cite{physics/9711021v2}.
$C$ is the value that minimizes $\textrm{MRF} \left(c \right)$.
In \cref{fig:mrf}, the projected number of remaining events based on the simulated background events as a function of $c$ is depicted.
At $C=0.9997$, at most, 10 background events with a mean value of 2 are projected to remain after applying the analysis to the full, statistically unblinded dataset.

\section{Systematics}
Some modeled parameters in simulation are associated with uncertainties which can result in systematic shifts of the models from reality.
Four systematic shifts for signal simulation have been investigated: a $\pm10\,\%$ shift of the DOM light detection efficiency~\cite{Aartsen:2016nxy}, variations of the angular acceptance of light of the DOMs~\cite{MSU_HoleIce}, correlated $\pm5\,\%$ variations in the scattering and absorption of the ice~\cite{aartsen2013measurement}, and $\pm 40\,\%$ variations in the effective luminescence light yield~\cite{pollmanicrc2021}.
Specialised simulation is conducted for each effect and the velocity dependent lowest signal retention for each effect are combined for the total systematic shift depicted in \cref{fig:sys}.
An expected shift at low velocities can be seen which is expected as events in this range are already dimmer and thus more susceptible to further systematic loss of brightness.
A second shift can be seen in the transition region between luminescence light and indirect Cherenkov light at about 0.55\,c which is only present for the final event selection step.

\section{Results}
\begin{figure}
    \centering
    \includegraphics[width=\textwidth]{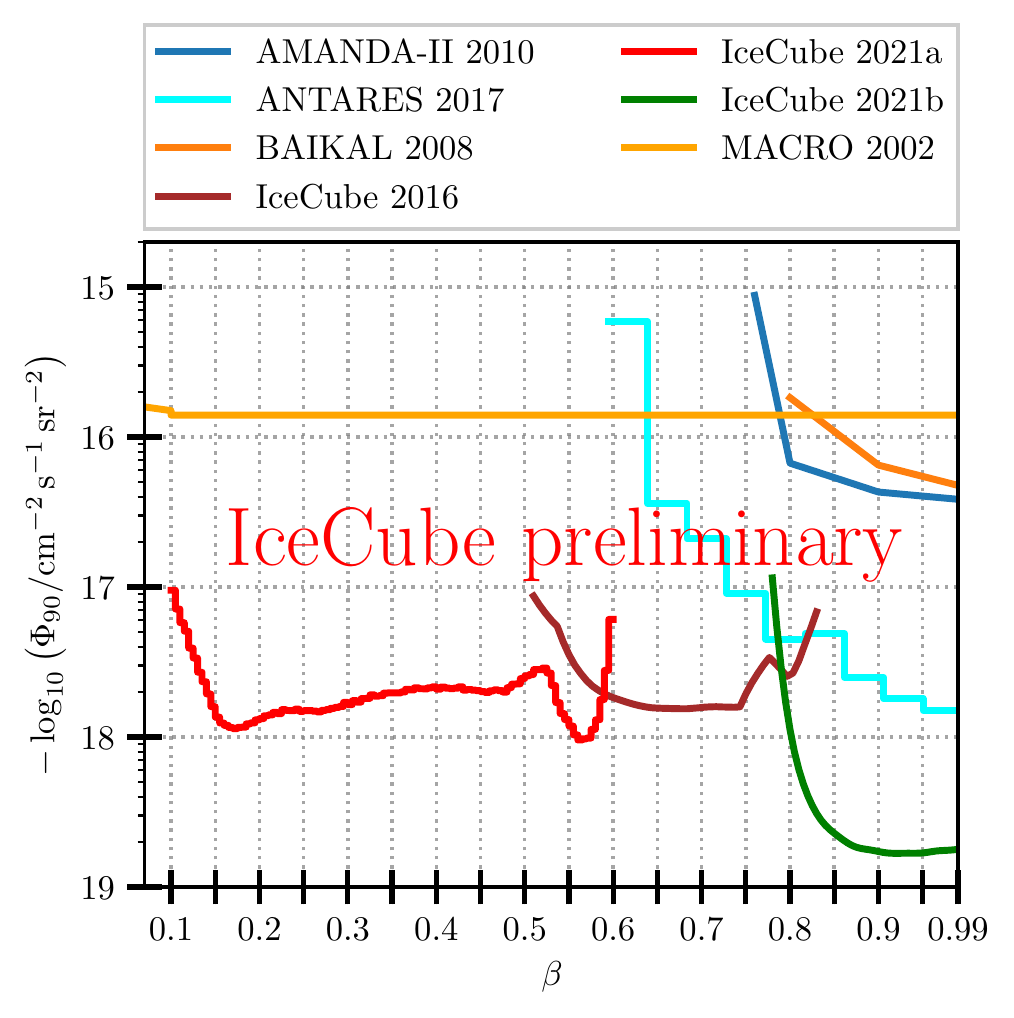}
    \caption{A chart with the flux limit presented in this contribution (IceCube 2021a) in contrast to previous searches ranging from the low relativistic to relativistic regime as a function of velocity is shown.
	Previous flux limits are taken from different experiments and collaborations, sorted in the legend by alphabetical order.
	The corresponding references, in the same order, are 
	\cite{Abbasi2010, 1703.00424v2, aynutdinov2008search, 1511.01350v2, burgmann2021, hepex0207020v2}.}
    \label{fig:flux_limit}
\end{figure}

The analysis has been applied to $2524.6$\,days of measured data taken at IceCube.
Two candidate events remained which is compatible with the expected number of remaining background events.
An upper limit on the flux of magnetic monopoles has been derived superseding previous best flux limits in the low relativistic regime by two orders of magnitude.
The derived flux limit in contrast to other searches for magnetic monopoles is illustrated in \cref{fig:flux_limit}.

\bibliographystyle{ICRC}
\bibliography{references}

\providecommand{\href}[2]{#2}\begingroup\raggedright\begin{thebibliography}{10}

\bibitem{Aartsen:2016nxy}
{\bfseries IceCube} Collaboration, M.~G. Aartsen {\em et~al.}
  \href{http://dx.doi.org/10.1088/1748-0221/12/03/p03012}{{\em JIONAS}
  {\bfseries 12} no.~03, (2017) 03012}.

\bibitem{Aartsen2014}
M.~G. Aartsen {\em et~al.}
  \href{http://dx.doi.org/10.1140/epjc/s10052-014-2938-8}{{\em EPJ C}
  {\bfseries 74} (2014) 2938}.

\bibitem{1511.01350v2}
{\bfseries IceCube} Collaboration, M.~G. Aartsen {\em et~al.}
  \href{http://dx.doi.org/10.1140/epjc/s10052-016-3953-8}{{\em EPJ C}
  {\bfseries 76} (2016) 133}.

\bibitem{Dirac:423077}
P.~A.~M. Dirac \href{http://dx.doi.org/10.1098/rspa.1931.0130}{{\em PRLAAZ}
  {\bfseries 133} (1931) 60--72}.

\bibitem{ahlen1978stopping}
S.~P. Ahlen \href{http://dx.doi.org/10.1103/PhysRevD.17.229}{{\em PRD}
  {\bfseries 17} no.~1, (1978) 229--233}.

\bibitem{Quickenden_1982}
T.~I. Quickenden {\em et~al.} \href{http://dx.doi.org/10.1063/1.444352}{{\em
  JCPSA6} {\bfseries 77} no.~8, (1982) 3790--3802}.

\bibitem{Freeman_1984}
C.~G. Freeman {\em et~al.} \href{http://dx.doi.org/10.1063/1.447691}{{\em
  JCPSA6} {\bfseries 81} no.~12, (1984) 5252--5254}.

\bibitem{ScienceDirect:S016943320801369X}
C.~Lee {\em et~al.} \href{http://dx.doi.org/10.1016/j.apsusc.2008.05.344}{{\em
  ASUSEE} {\bfseries 255} no.~9, (2009) 4716--4719}.

\bibitem{quickenden1991effect}
T.~I. Quickenden {\em et~al.} \href{http://dx.doi.org/10.1063/1.461217}{{\em
  JCPSA6} {\bfseries 95} no.~12, (1991) 8843--8852}.

\bibitem{ScienceDirect:135901978890166X}
M.~G. Bakker {\em et~al.}
  \href{http://dx.doi.org/10.1016/1359-0197(88)90166-X}{{\em RPCHDM} {\bfseries
  32} no.~6, (1988) 767--772}.

\bibitem{pollmann2019enabling}
{\bfseries IceCube} Collaboration, A.~M. Pollmann
  \href{http://dx.doi.org/10.22323/1.358.0983}{{\em PoS(ICRC 2019)} {\bfseries
  358} (2019) 983}.

\bibitem{Tanabashi:2636832}
{\bfseries Particle Data Group} Collaboration, M.~Tanabashi {\em et~al.}
  \href{http://dx.doi.org/10.1103/PhysRevD.98.030001}{{\em PRD} {\bfseries 98}
  no.~3, (2018) 030001}.

\bibitem{heck1998corsika}
D.~Heck {\em et~al.}, ``{CORSIKA}: {A} {Monte Carlo} code to simulate extensive
  air showers,'' 1998.

\bibitem{gaisser2011spectrum}
T.~K. Gaisser \href{http://dx.doi.org/10.1016/j.astropartphys.2012.02.010}{{\em
  APHYEE} {\bfseries 35} no.~12, (2012) 801--806}.

\bibitem{astro-ph/0611418v3}
M.~Honda {\em et~al.} \href{http://dx.doi.org/10.1103/PhysRevD.75.043006}{{\em
  PRD} {\bfseries 75} no.~4, (2007) 043006}.

\bibitem{aartsen2014improvement}
{\bfseries IceCube} Collaboration, M.~G. Aartsen {\em et~al.}
  \href{http://dx.doi.org/10.1016/j.nima.2013.10.074}{{\em NIMAER} {\bfseries
  736} (2014) 143--149}.

\bibitem{Chen:2016:XST:2939672.2939785}
T.~Chen and C.~Guestrin \href{http://dx.doi.org/10.1145/2939672.2939785}{{\em
  KDD '16} (2016) 785--794}.

\bibitem{breiman1996bagging}
L.~Breiman \href{http://dx.doi.org/10.1007/BF00058655}{{\em Machine Learning}
  {\bfseries 24} no.~2, (1996) 123--140}.

\bibitem{physics/9711021v2}
G.~J. Feldman and R.~D. Cousins
  \href{http://dx.doi.org/10.1103/PhysRevD.57.3873}{{\em PRD} {\bfseries 57}
  no.~7, (1998) 3873--3889}.

\bibitem{MSU_HoleIce}
D.~Chirkin, ``Flasher data-derived ice models,'' tech. rep., IceCube
  Colloboration, 2017.
\newblock
  \url{https://docushare.icecube.wisc.edu/dsweb/Get/Document-79091/ice.pdf}.

\bibitem{aartsen2013measurement}
{\bfseries IceCube} Collaboration, M.~G. Aartsen {\em et~al.}
  \href{http://dx.doi.org/10.1016/j.nima.2013.01.054}{{\em NIMAER} {\bfseries
  711} (2013) 73--89}.

\bibitem{pollmanicrc2021}
{\bfseries IceCube} Collaboration, A.~M. Pollmann {\em PoS(ICRC 2021) (these
  proceedings)} 1093.

\bibitem{Abbasi2010}
{\bfseries IceCube} Collaboration, R.~Abbasi {\em et~al.}
  \href{http://dx.doi.org/10.1140/epjc/s10052-010-1411-6}{{\em EPJ C}
  {\bfseries 69} no.~3, (2010) 361--378}.

\bibitem{1703.00424v2}
{\bfseries ANTARES} Collaboration, A.~Albert {\em et~al.}
  \href{http://dx.doi.org/10.1007/JHEP07(2017)054}{{\em JHEP} {\bfseries 2017}
  no.~7, 54}.

\bibitem{aynutdinov2008search}
V.~Aynutdinov {\em et~al.}
  \href{http://dx.doi.org/10.1016/j.astropartphys.2008.03.006}{{\em APHYEE}
  {\bfseries 29} no.~6, (2008) 366--372}.

\bibitem{burgmann2021}
{\bfseries IceCube} Collaboration, A.~M. Pollmann {\em VLVnT 2021} (2021) .

\bibitem{hepex0207020v2}
{\bfseries MACRO} Collaboration, M.~Ambrosio {\em et~al.}
  \href{http://dx.doi.org/10.1140/epjc/s2002-01046-9}{{\em EPJ C} {\bfseries
  25} (2002) 511--522}.

\end{thebibliography}\endgroup
\end{document}